\documentclass [12pt]{iopart}
\usepackage {iopams}
\usepackage{setstack}
\usepackage{amstext}
\usepackage{amsfonts}
\usepackage{graphicx}
\usepackage{subfigure}
\usepackage{array}
\usepackage{appendix}
\usepackage{mathrsfs}
\usepackage{color}
\usepackage{makecell}
\usepackage{adjustbox,lipsum}
\usepackage{upgreek}
\usepackage{algorithm}
\usepackage{algpseudocode}
\usepackage{makecell}
\usepackage{adjustbox,lipsum}

\begin{document}
\title[]{Quantum superposing algorithm for quantum encoding}

\author{Jaehee~Kim$^{1}$, Taewan~Kim$^{2}$, Kyunghyun~Baek$^{2}$, Yongsoo~Hwang$^{2}$, Joonsuk~Huh$^{1,3,4}$, and Jeongho~Bang$^{2}$}

\address{$^1$ Sungkyunkwan University Advanced Institute of Nanotechnology, Suwon, 16419, Korea}
\address{$^2$ Electronics and Telecommunications Research Institute, Daejeon, 34129, Korea}
\address{$^3$ Department of Chemistry, Sungkyunkwan University Suwon, 16419, Korea}
\address{$^4$ Institute of Quantum Biophysics, Sungkyunkwan University, Suwon 16419, Korea}

\vspace{10pt}

\begin{indented}
\item The first two authors (J.K. and T.K.) contributed equally to this study. 
\item Correspondence and requests for materials should be addressed to J.H. and J.B.
\end{indented}

\ead{\mailto{joonsukhuh@skku.edu} and \mailto{jbang@etri.re.kr}}

\begin{abstract}
Efficient encoding of classical data into quantum state---currently referred to as quantum encoding---holds crucial significance in quantum computation. For finite-size databases and qubit registers, a common strategy of the quantum encoding entails establishing a classical mapping that correlates machine-recognizable data addresses with qubit indices that are subsequently superposed. Herein, the most imperative lies in casting an algorithm for generating the superposition of any given number of qubit indices. This algorithm is formally known as quantum superposing algorithm. In this work, we present an efficient quantum superposing algorithm, affirming its effectiveness and superior computational performance in a practical quantum encoding scenario. Our theoretical and numerical analyses demonstrate a substantial enhancement in computational efficiency compared to existing algorithms. Notably, our algorithm has a maximum of $2n - 3$ controlled-not (CNOT) counts, representing the most optimized result to date.
\end{abstract}

\maketitle

\newtheorem{result}{Result}
\newcommand{\bra}[1]{\left<#1\right|}
\newcommand{\ket}[1]{\left|#1\right>}
\newcommand{\abs}[1]{\left|#1\right|}
\newcommand{\norm}[1]{\left|\!\left| #1\right|\!\right|}
\newcommand{\expt}[1]{\left<#1\right>}
\newcommand{\braket}[2]{\left<{#1}|{#2}\right>}
\newcommand{\commt}[2]{\left[{#1},{#2}\right]}
\newcommand{\round}[1]{\ensuremath{\lfloor#1\rceil}}

\newcommand{\identity}{1\!\!1}

\section{Introduction}

The primary challenge for developing quantum computing algorithms lies in establishing a strong theoretical proof of quantum computational speedup~\cite{Montanaro2016}. However, the absence of a robust assumption regarding quantum superposed data hinders the claim of realizing the proved quantum computational speedup, necessitating careful consideration~\cite{Aaronson2015, Tang2021, Cotler2021}. Thus, contemporary focus extends beyond the theoretical proofs to evaluating the practical feasibility of the demonstrated quantum speedups throughout the entire computing process. In particular, the topic recently emerged as significant is, ``How can we efficiently prepare a quantum superposition of a substantial volume of data?" or equivalently, ``How can the computational overhead associated with data superposition be optimized without compromising the speedup achieved in the main computation?"~\cite{Weigold2021, Araujo2021, Mozafari2021, Zhang2022}. This issue is of particular significance in the field of quantum machine learning (QML)~\cite{Biamonte2017, Bang2019, Havlivcek2019, Zoufal2019, Cerezo2022}. 

In a general context, data is presented in a user-friendly format, comprising quantities that are inherently incomputable. However, for machine access, the data requires translation into machine-recognizable addresses. The conventional assumption posits that each machine-recognizable address is mapped to a bit string~\cite{Blakeley1996}. In quantum computation, these bit strings must undergo the quantum superposition to utilize quantum parallelism effectively. Facilitating this process necessitates a quantum superposing algorithm designed to create a quantum superposition of bit strings. The development and implementation of such an efficient algorithm is computationally hard, because it must consistently operate for an arbitrary number of data. Distinguishing this problem from the simple (textbook) example of creating $N=2^n$ superposition states with $n$ Hadamard gates is important (see CASE I in Table~\ref{tab:N_CNOTs})~\cite{Mozafari2021, Mozafari2022, De2022, Araujo2023}\footnote{One might consider achieving a $2^n$ superposition state using additional qubit registers and dummy data with the Hadamard gates. However, this approach does not contribute effectively to overall quantum computing tasks, particularly in the context of quantum data access and/or encoding (for example, see Ref.~\cite{Shukla2024-2}).}.

In this work, our focus is on the development of an efficient quantum superposing algorithm tailored for a practical quantum data encoding scenario. To elucidate the problem, let us consider a scenario of encoding an arbitrarily given number, say $N$, of the bit strings into a quantum state~\cite{Plesch2011}
\begin{eqnarray}
\ket{\Psi} = \sum_{\mathbf{j} \in B} \alpha_\mathbf{j} \ket{\mathbf{j}},
\label{eq:sup_general}
\end{eqnarray}
where $\alpha_\mathbf{j}$ are quantum amplitudes, $\mathbf{j}$ denotes an $n$-bit string, i.e., $\mathbf{j}=j_0 j_1 \cdots j_{n-1}$ ($j_k \in \{0, 1\}$), and $n = \lceil \log_2{N} \rceil$, which is smaller than or equal to the total size of the qubit register in an implementing quantum machine. The set $B$ comprises $\mathbf{j}$'s mapped to the addresses of actual data, with $\abs{B}$ denoting the cardinality of $B$, i.e., $N$. Here, we set $\alpha_\mathbf{j}=\frac{1}{\sqrt{N}}$ for all $\mathbf{j}$ corresponding to the bit strings of the consecutive numbers from 0 to $N-1$. This problem setting is commonly encountered, for example, in quantum data search~\cite{Grover1997, Brassard2002, Viamontes2005, Broda2016} and various QML tasks~\cite{Biamonte2017, Havlivcek2019, Zoufal2019}. In such a scenario, we present an efficient quantum superposing algorithm. Our presented algorithm exhibits a controlled-not (CNOT) complexity of $O(n)$, specifically with a maximum of $2n - 3$ CNOT counts, which is the most optimized results so far\footnote{Notably, a recently proposed algorithm achieves a $T$-count of $O(n + \log{\frac{1}{\epsilon}})$ for a fault-tolerant quantum computation with a success probability of $1-\epsilon$; however, an unavoidable error $\epsilon$ and an approximate execution mode (see Section III-D in Ref.\cite{Babbush2018})}~\cite{Yu2017,Shukla2024}. Such a reduction in computational cost is attributed to algorithmic optimizations, without relying on multi-controlled gates such as $n$-Toffoli gates~\cite{Mottonen2005}. The superior efficiency of our algorithm is substantiated through rigorous theoretical and numerical analyses, promising an overall enhancement in quantum encoding efficiency. 

\section{Data encoding and superposing algorithm}

\subsection{Encoding data into equal superposition} 

Encoding refers to the process of mapping data or information into specific patterns. In quantum computation, this entails mapping data into a quantum state, such that\footnote{The upper tilde denotes ``data structure,'' which should be distinguished from ``set.''}:
\begin{eqnarray}
\text{data}: \tilde{D} \to \text{data-encoded quantum state}: \ket{\psi(\tilde{D})}.
\end{eqnarray}
This process is currently referred to as ``quantum encoding"~\cite{Schuld2015,Schuld2018book,Weigold2021,Schuld2021}. While various quantum encoding methods exist, here we present a framework specifically designed for generating an equal superposition of data, with a clear distinction between classical and quantum components. The classical aspect handles the user-intuitive data access, assuming the full utilization of existing classical data-processing techniques. Meanwhile, the quantum component is responsible for producing a quantum superposed state with finite-sized, say $\Xi$, qubit registers and a quantum superposing algorithm.

\begin{figure}[t]
 \centering
 \includegraphics[width=0.65\textwidth]{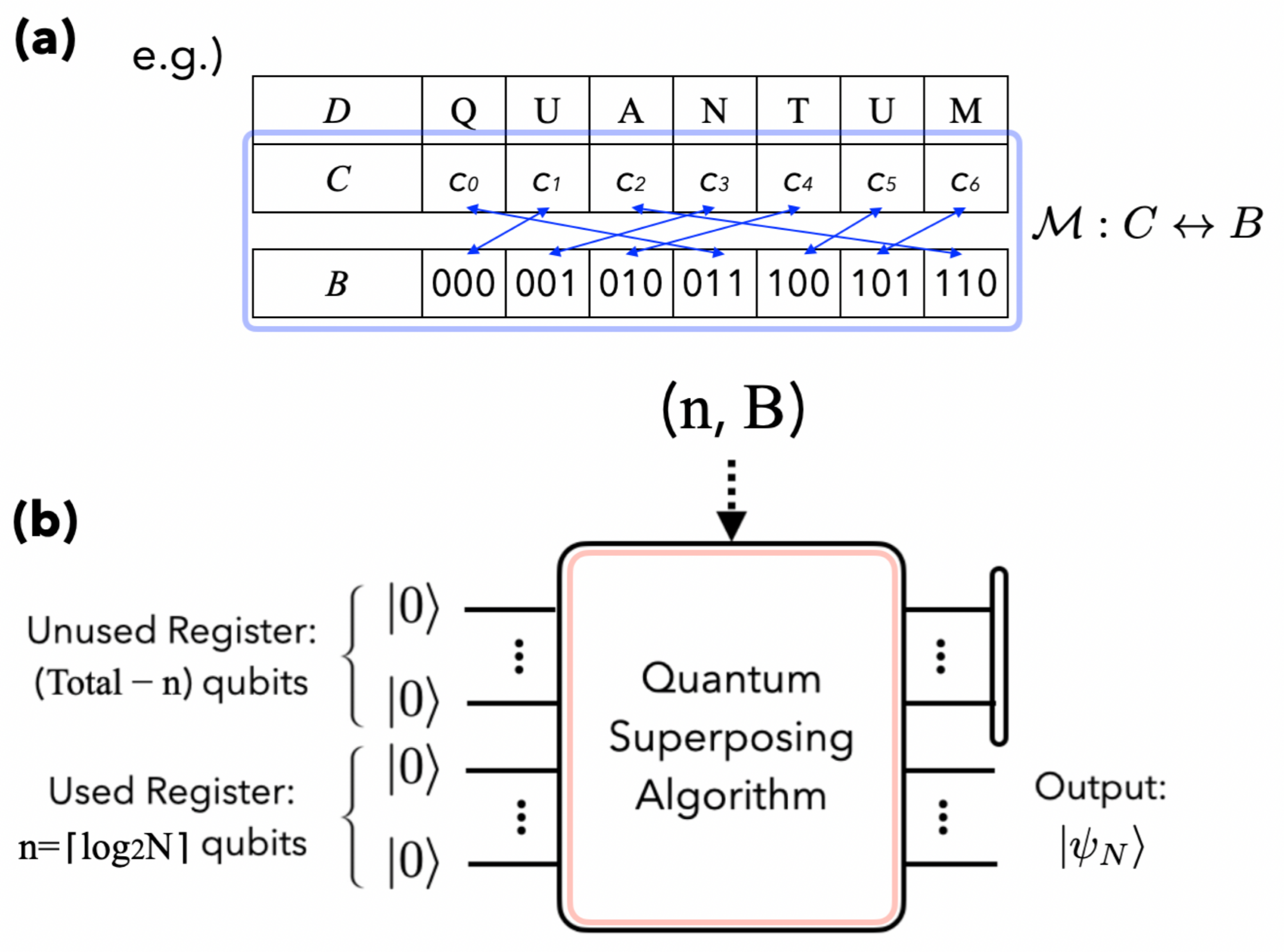}
 \caption{\label{fig:QDSscheme} Data encoding schematic. (a) Given a dataset of $N$ data items, a set $B$ of the sequential binary indices is first defined. Here, $\abs{B}=N$ and $n = \lceil \log_2{N} \rceil$. This construction leads to a mapping ${\cal M}$ that ``arbitrarily" connects the address of the allocated data with the indices in $B$. (b) Subsequently, a superposing algorithm is employed to prepare a superposition state involving the indices extracted from $B$. To illustrate this process, a straightforward example using the data $\tilde{D}=[ \text{Q} | \text{U} | \text{A} | \text{N} | \text{T} | \text{U} | \text{M} ]$ is depicted in (a). For more details, refer to the main text.}
\end{figure}

In this framework, the quantum encoding steps are as follows: (i) Start with a data $\tilde{D}$ of size $N$, where each element has an intrinsic and classical-accessible address $c \in C(\tilde{D})$. Here, $C(\tilde{D})$ denotes a set of the addresses mediating the data $\tilde{D}$. (ii) Create a set $B$ with the number $N$ of bit-string indices, mapping each index to an address $c$. While there are $N!$ potential mappings, only one such mapping is selected {\em arbitrarily}. This mapping, denoted as ${\cal M}$, is one-to-one and bidirectional: i.e., 
\begin{eqnarray}
{\cal M}: c \in C(\tilde{D}) \leftrightarrow b \in B,
\end{eqnarray}
where $b$ denotes a binary element of $B$. Here, we note that ${\cal M}$ does not function as a conventional look-up-table (LUT) that is commonly utilized to alleviate the computational load by retrieving the pre-computed results. ${\cal M}$ is maintained until the main quantum computing task is completed. These two steps (i)-(ii) are managed by the classical system. Subsequently, ($n$, $B$) is delivered to the quantum part, where $n = \lceil \log_2{N} \rceil$. (iii) In the quantum component, the quantum superposing algorithm is executed to generate a superposition state of the bit-string indices in $B$. Thus, our data encoding scheme can be described as follows:
\begin{eqnarray}
\tilde{D} \to \left( \frac{1}{\sqrt{N}} \sum_{b \in B} \ket{b}, {\cal M} \right).
\end{eqnarray}
The efficiency of this scheme can be analyzed based on the computational complexity of the quantum superposing algorithm (see Fig.~\ref{fig:QDSscheme}). To understand better, let us consider a data $\tilde{D}$ with $N=7$, such as
\begin{eqnarray}
\tilde{D}=[ \text{Q} | \text{U} | \text{A} | \text{N} | \text{T} | \text{U} | \text{M} ].
\end{eqnarray}
In the classical system, these alphabet characters can be allocated with the addresses:
\begin{eqnarray}
\tilde{C}=[ c_0 | c_1 | c_2 | c_3 | c_4 | c_5 | c_6 ].
\end{eqnarray}
Accordingly, we can set up the bit-string indices:
\begin{eqnarray}
B= \{ 000, 001, 010, 011, 100, 101, 110 \}
\end{eqnarray}
with $n=\lceil \log_2{7} \rceil = 3$. The mapping ${\cal M}$ between $B$ and $C(\tilde{D})$ is formed, {\em arbitrarily}, such that
\begin{eqnarray}
\fl ~~~~(\text{`000'}:c_1), (\text{`001'}:c_3), (\text{`010'}:c_4), (\text{`011'}:c_0), (\text{`100'}:c_5), (\text{`101'}:c_6), (\text{`110'}:c_2).
\end{eqnarray}
Using this setup, the quantum superposing algorithm generates the quantum state as
\begin{eqnarray}
\frac{1}{\sqrt{7}} \left( \ket{000} + \ket{001} + \ket{010} + \ket{011} + \ket{100} + \ket{101} + \ket{110}\right).
\end{eqnarray}

This framework has been implicitly utilized and offers advantages through a distinct separation between classical and quantum components. Specifically, it affords flexibility in utilizing data based on computational requirements. For tasks efficiently handled by classical computers alone, computations can proceed using steps (i) and (ii). When quantum computing is required, step (iii) can be invoked to activate the quantum superposing algorithm. This perspective aligns well with classical-quantum hybrid computing. Notably, in this scenario, accessing the actual data during quantum computation is unnecessary, and it can enhance both data integrity~\cite{Song2021a} and security~\cite{Song2021b, Harney2022}.

\subsection{Quantum superposing algorithm}  

Now, we present our quantum superposing algorithm. Diverging from previous approaches employing resource-intensive multi-controlled gates, our algorithm incorporates only four types of quantum gates: Hadamard $\hat{H}$, zero-controlled-$\hat{H}$ (denoted ``zero-c$\hat{H}$''), $\hat{G}(p)$, controlled-$\hat{G}(p)$ (denoted ``c$\hat{G}(p)$''). The single-qubit Hadamard $\hat{H}$ creates the superposition state $\frac{1}{\sqrt{2}}\left(\ket{0}+(-1)^j \ket{1}\right)$ for the input $\ket{j}$. The zero-c$\hat{H}$ performs the Hadamard $\hat{H}$ on the target qubit if the control qubit is in the state $\ket{0}$. Here, $\hat{G}(p)$ is defined as follows (for $0\le p \le 1$)~\cite{Yoo2014,Lee2019,Mozafari2021}:
\begin{eqnarray}
\hat{G}(p) = \hat{R}_{y}\left(2\cos^{-1}(\sqrt{p})\right)
	=  \left(
	\begin{array}{cc}
		    \sqrt{p} & -\sqrt{1-p}\\
		    \sqrt{1-p} & \sqrt{p}
	\end{array}
	\right),
\end{eqnarray}
where $\hat{R}_y(\theta)$ is a $y$-rotating operation with the angle of $\theta$. We note that $\hat{G}(\frac{1}{2})\ket{0}=\hat{H}\ket{0}$, and $\hat{G}(\frac{1}{2})$ is equivalent to $\hat{H}$ for the input $\ket{0}$. The c$\hat{G}(p)$ performs $\hat{G}(p)$ on the target qubit conditioned on the control qubit being in the state $\ket{1}$.

Our algorithm starts with $N$ as input. As the output, the algorithm provides the quantum circuit for preparing the state,
\begin{eqnarray}
\ket{\psi_N} = \frac{1}{\sqrt{N}}\sum_{\mathbf{j} \in B} \ket{\mathbf{j}},
\end{eqnarray}
where $\mathbf{j}=j_0 j_1 \cdots j_{n-1}$ ($j_k \in \{0, 1\}$) is the sequential binary indices from $00\ldots0$ and $|B|=N$. The overall process of our algorithm is as follows (see $\mathbf{Algorithm~\ref{alg1:pse_QS}}$). Given $n=\lceil \log_2(N)\rceil $, the qubit register to be used is assigned as $q[0], q[1], \cdots, q[n-1]$. The state of the register $q[k]$ is initialized to $\ket{0}$ for all $k=0,1,\ldots,n-1$. Here, note that a given arbitrary $N$ can be represented as $2^{\xi}M$, where $0 \le \xi \le n$ and $M$ is odd. With such a representation $N=2^\xi M$, the number $\xi$ of Hadamard gates, i.e., $\hat{H}^{\otimes \xi}$, can be employed to generate a superposition,
\begin{eqnarray}
\frac{1}{\sqrt{2^\xi}}\left( \ket{0}+\ket{1} \right)^{\otimes \xi}.
\end{eqnarray}
We apply $\hat{H}^{\otimes \xi}$ to the registers from $q[n-\xi]$ to $q[n-1]$. The remaining task is to prepare the state, named the ``odd-number superposed state", as follows (for $M\neq1$):
\begin{eqnarray}
\frac{1}{\sqrt{M}}\sum_{\mathbf{j} \in B'} \ket{\mathbf{j}},
\label{eq:odd_superp}
\end{eqnarray}
where $\mathbf{j}=j_0 j_1 \cdots j_{n-\xi-1}$ ($j_i \in \{0, 1\}$) refers to the binary indices from $0\ldots0$ and $\abs{B'}=M$. This state is generated in the registers $q[0], q[1], \ldots, q[n-\xi-1]$. The sub-process for preparing the odd-number superposed state is delineated in $\mathbf{Algorithm~\ref{alg2:pse_QS_odd}}$. The majority of the computational cost of the algorithm is concentrated in this sub-process, rendering $\mathbf{Algorithm~\ref{alg2:pse_QS_odd}}$ the central component of the overall algorithm. The algorithm is completed by combining the states in all qubit registers, such that 
\begin{eqnarray}
\ket{\psi_N} = \frac{1}{\sqrt{M}}\sum_{\mathbf{j} \in B'} \ket{\mathbf{j}} \otimes \frac{1}{\sqrt{2^\xi}}\left( \ket{0}+\ket{1} \right)^{\otimes \xi}.
\end{eqnarray}
\begin{algorithm}[H]
	\caption{Overall frame}
    \label{alg1:pse_QS}
    \hspace*{\algorithmicindent} \textbf{Input :} $N$ \\
    \hspace*{\algorithmicindent} \textbf{Output :} Circuit for preparing $\frac{1}{\sqrt{N}}\sum_{\mathbf{j} \in B} \ket{\mathbf{j}}$
	\begin{algorithmic}[1]
	\State{Calculate $n=\lceil \log_2{N} \rceil$}
        \State{Factor $N$ as $2^\xi M$ ($0 \le \xi \le n$ and $M$ is odd)}
        \State{Assign the qubit registers $q[0], q[1], \cdots, q[n-1]$.}
        \State{Initialize all qubit registers to be in $\ket{0}$.}
        \If{$\xi \geq 1$}
            \For{$k \gets 1$ \textbf{to} $\xi$}                                       
                \State {Apply $\hat{H}_{q[n-k]}$}
            \EndFor
        \EndIf
        \If{$\xi = 0$} 
            \State{Go to $\mathbf{Algorithm~\ref{alg2:pse_QS_odd}}$.}
        \EndIf
	\end{algorithmic} 
	\Comment{\footnotesize ``$\hat{A}_{q[\cdot]}$'' denotes a gate $\hat{A}$ applied in $q[\cdot]$}
\end{algorithm} 

Now, we delve into the details of $\mathbf{Algorithm~\ref{alg2:pse_QS_odd}}$. This process is constructed using three types of gates: $\hat{G}(p)$, c$\hat{G}(p)$, and zero-c$\hat{H}$ gates; the single-qubit Hadamard gate $\hat{H}$ is dispensable. The odd number $M$ is represented in the following form:
\begin{eqnarray}
M=2^{k_0}+ 2^{k_1}+ \cdots+ 2^{k_{g-2}}+1,
\label{eq:odd_M}
\end{eqnarray}
where $k_0 > k_1 > \cdots > k_{g-2} > 0$ and $m=\lceil \log_2{M}\rceil$. The gates, i.e., $\hat{G}(p)$, c$\hat{G}(p)$, and zero-c$\hat{H}$, are applied to the registers $q[0], q[1], \ldots, q[n-\xi-1]$. Firstly, the single-qubit gate $\hat{G}(p_0)$ is applied to the register $q[0]$, where the parameter $p_0$ is determined as $p_0 = \frac{2^{k_0}}{M}$. Following this, the c$\hat{G}(p_j)$ gates are implemented. Herein, the parameters, $p_j$ ($1 \le j \le g-2$), are set as
\begin{eqnarray}
p_j = 2^{k_j} \left( M - \sum^{j-1}_{l=1} 2^{k_l} \right)^{-1},
\end{eqnarray}
and the control and target qubits are selected by the following rule:
\begin{eqnarray}
\text{c$\hat{G}(p_j)$}~\left\{
\begin{array}{ll}
\text{Control:} & q[m - k_{j-1} - 1], \\
\text{Target:} & q[m - k_j - 1],
\end{array}
\right.
\end{eqnarray}
with the value of $k_j$ being derived from Eq.~(\ref{eq:odd_M}). Finally, the zero-c$\hat{H}$ gates are applied as follows:
(i) Initially, zero-c$\hat{H}$ gates with $q[m - k_{g-2} - 1]$ as the control qubit and $q[m - k_{g-2} + j]$ as the target qubit are successively applied, incrementing $j$ from $0$ to $k_{g-2} - 1$.
(ii) Subsequently, zero-c$\hat{H}$ gates with the control qubit $q[m - k_{j-1} -1]$ and target qubit $q[m - k_j - l]$ are incorporated into the algorithm circuit, increasing $l$ from $1$ to $k_{j-1} - k_j$. This step (ii) is iteratively performed, commencing with $j=g-2$ and progressing down to $j=1$.
\begin{algorithm}[H]
	\caption{Process for generating the odd-number superposed state}
    \label{alg2:pse_QS_odd}
    \hspace*{\algorithmicindent} \textbf{Input :} $M$ \\
    \hspace*{\algorithmicindent} \textbf{Output :} Circuit for preparing $\frac{1}{\sqrt{M}}\sum_{\mathbf{j} \in B'} \ket{\mathbf{j}}$
	\begin{algorithmic}[1]
	\State{Calculate $m=\lceil \log_2{M} \rceil$}
        \State{Express as $M=2^{k_0}+ 2^{k_1}+ \cdots+ 2^{k_{g-2}}+1$ ($k_0>k_1>\ldots>k_{g-2}>0$)}
        \State{Apply $\hat{G}(p_{0})_{q[0]}$ with $p_{0}=\frac{2^{k_0}}{M}$}
        \For{$i \gets 1$ \textbf{to} $g-2$}
            \State{$\beta_c = m-k_{i-1}-1$ and $\beta_t=m-k_{i}-1$} 
            \State{Apply $c\hat{G}(p_i)_{q[\beta_c], q[\beta_t]}$ with $p_{i}=\frac{2^{k_i}}{M-\sum^{i-1}_{j=1} 2^{k_j}}$}
        \EndFor
        \For{$j \gets 0$ \textbf{to} $k_{g-2}-1$}
            \State{$\beta_c = m - k_{g-2} - 1$ and $\beta_t = m - k_{g-2} + j$}
            \State{Apply zero-c$\hat{H}_{q[\beta_c], q[\beta_t]}$} 
        \EndFor
        \For{$j \gets g-2$ \textbf{downto} $1$}
            \For{$l \gets 1$ \textbf{to} $k_{j-1} - k_{j}$} 
            \State{$\beta_c = m - k_{j-1} - 1$ and $\beta_t = m - k_{j} - l$}
            \State{Apply zero-c$\hat{H}_{q[\beta_c], q[\beta_t]}$} 
            \EndFor
        \EndFor
	\end{algorithmic}
	\Comment{\footnotesize ``$q[\beta_c]$'' and ``$q[\beta_t]$'' denote the control and target qubits.}
\end{algorithm}

\section{Analysis}

\begin{figure}[t]
	\centering
	\includegraphics[width=0.70\textwidth]{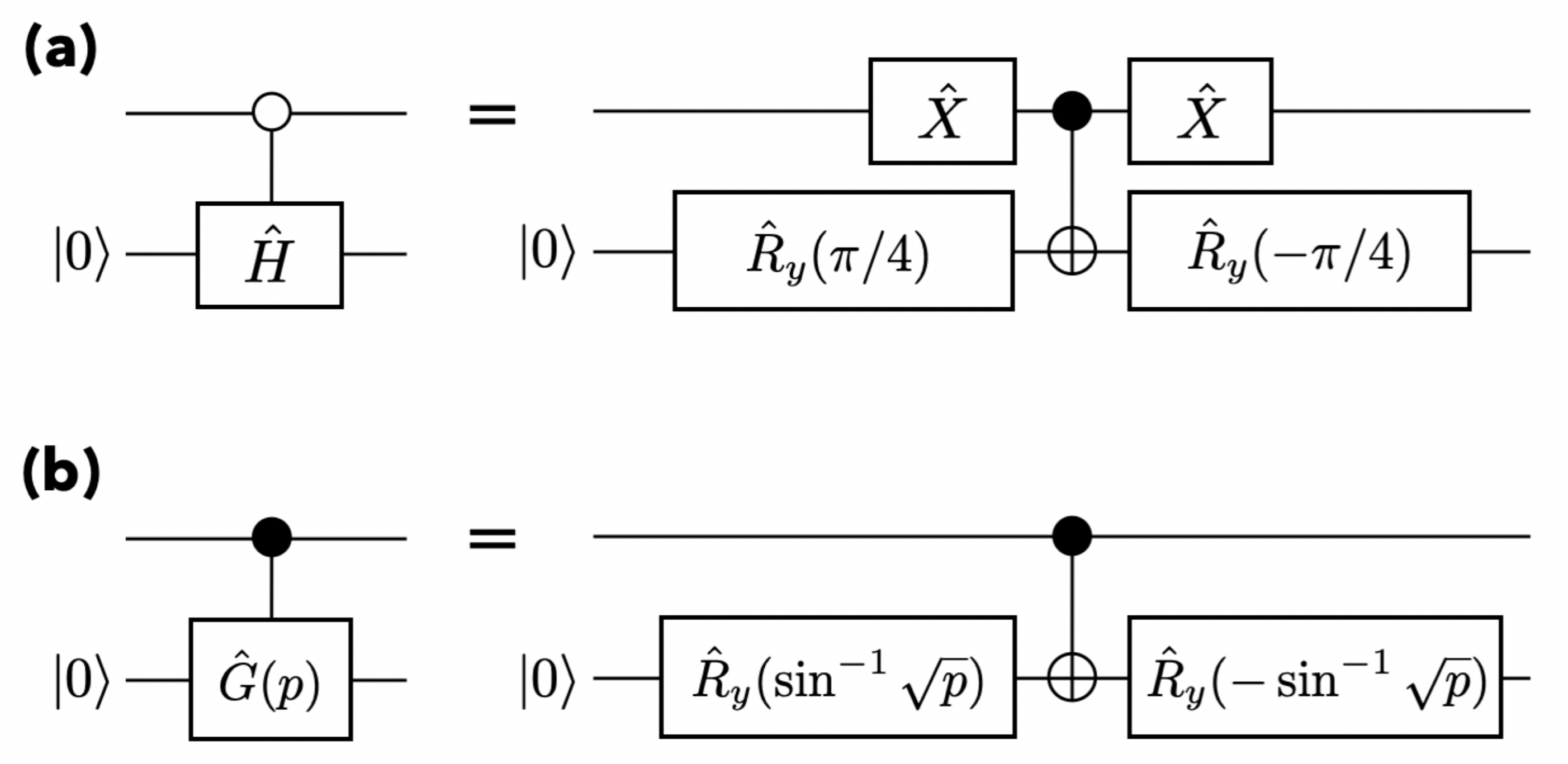}
	\caption{Decompositions of the two-qubit gates: (a) zero-c$\hat{H}$ and (b) c$\hat{G}(p)$. The decompositions use {\em only one CNOT gate} with the input of the target qubit being $\ket{0}$.}
	\label{fig:z-cH,CG}
\end{figure}

The CNOT gate plays a crucial role as one of the universal gates for entanglement creation, yet its ready availability and reliability in a quantum circuit are not guaranteed. Moreover, CNOT gates are more error-prone than other single-qubit gates. For these reasons, counting the number of CNOT gates has become a valuable metric for assessing the efficiency of quantum circuits or algorithms, referred to as ``CNOT complexity"~\cite{Shende2004, Shende2005, Plesch2011}. In our analysis, we thus employ CNOT complexity to evaluate the performance of our quantum superposing algorithm.

Our algorithm primarily utilizes two distinct two-qubit gates: the zero-c$\hat{H}$ gate and c$\hat{G}(p)$ gate. To start the analysis, we present an efficient gate decomposition where both the zero-c$\hat{H}$ gate and c$\hat{G}(p)$ gate are realized using only one CNOT gate, as illustrated in Fig.~\ref{fig:z-cH,CG}(a) and~\ref{fig:z-cH,CG}(b). Executing either the zero-c$\hat{H}$ gate or c$\hat{G}(p)$ can thus be treated as a single use of CNOT. Here, it should be indicated that this c$\hat{G}(p)$ gate implementation, using only a single CNOT, is effective when the target input is in $\ket{0}$ state. While this implementation is not universally applicable, it is the most optimized for our algorithm. Actually, it contributes to a reduction in the overall number of required CNOT gates\footnote{This results in a reduction of the required number of CNOT gates by up to approximately 50\% in the case of $c\hat{G}$ compared with the previous method.}. Therefore, the CNOT complexity of our algorithm is contingent on $\mathbf{Algorithm~\ref{alg2:pse_QS_odd}}$. Within our algorithm, a maximum of $n-2$ c$\hat{G}(p)$ gates and a maximum of $n-1$ zero-c$\hat{H}$ gates can be applied, as evident from the ``for loops" in $\mathbf{Algorithm~\ref{alg2:pse_QS_odd}}$. 

\begin{table}[t]
\centering
\setlength{\tabcolsep}{0.20in}
\renewcommand{\arraystretch}{1.2}
\begin{tabular}{c  c  c  c}
\hline\hline
CASE 	& $N = 2^\xi M$ 			& Employing Gates 					& \thead{Number of CNOTs \\ ($=g+m-3$ for $g\geq2$)} \\
\hline
I	& $N=2^n$ $(M=1)$  		& $\hat{H}$ 						& $0$ ($g=1$) \\ 
II 	& $N=2^{n-1} + 1$ $(N=M)$  	& $\hat{G}$, zero-c$\hat{H}$ 			& $n-1$ \\ 
III 	& $N=2^{n} - 1$ $(N=M)$  	& $\hat{G}$, c$\hat{G}$, zero-c$\hat{H}$ 	& $2n-3$ \\ 
IV 	& $\xi=0, M \neq 1$ $(N=M)$  	& $\hat{G}$, c$\hat{G}$, zero-c$\hat{H}$ 	& $\le 2n-4$ \\ 
V 	& $\xi \ge 1$ $(N = 2^\xi M)$ 	& $\hat{G}$, c$\hat{G}$, zero-c$\hat{H}$ 	& $\le 2n-5$ \\ 
\hline\hline
\end{tabular}
\caption{Given $N=2^\xi M$, we classify the five cases and analyze the number of CNOTs for each specific case (see the main text for details).}
\label{tab:N_CNOTs}
\end{table}

\begin{table}[t]
\centering
\setlength{\tabcolsep}{0.17in}
\renewcommand{\arraystretch}{1.2}
\begin{tabular}{c  c}
\hline\hline
CASE 	& $N$ \\
\hline
I  		& $2,4,8,16^\star,\ldots$	  \\ 
II	  	& $3,5,9,17^\star,\ldots$  \\ 
III		& $7,15,31^\star,\ldots$  \\ 
IV		& $11,13,19,21,2 3,25,27,29^\star,\ldots$  \\ 
V		& $6,10,12,14,18,20,22,24,26,28,30^\star,\ldots$  \\ 
\hline\hline
\end{tabular}
\caption{Specific $N$ numbers corresponding to the cases in Table~\ref{tab:N_CNOTs}.}
\label{tab:N}
\end{table}

\begin{figure}[t]
	\centering
	\includegraphics[width=1.0\textwidth]{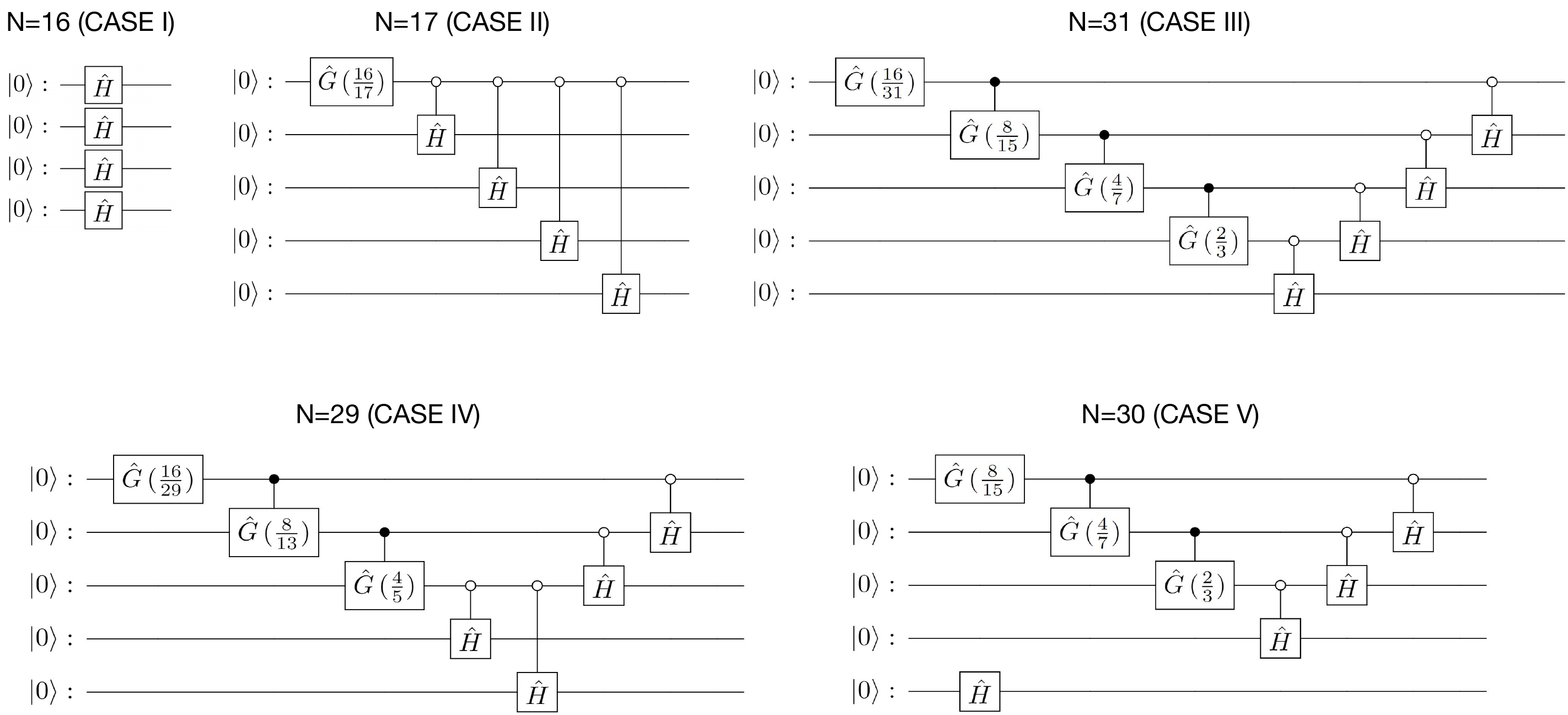}
	\caption{Examples of quantum circuits corresponding to each case classified in Table~\ref{tab:N_CNOTs}. The most demanding case is the third.}
	\label{fig:circuit_examples}
\end{figure}

Continuing the analysis, we generalize the form of $N$ in terms of $M$ based on the computational efficiency demonstrated by our algorithm, we generalize the types of $N$. Subsequently, we categorize different cases of $N$ and delve into detailed descriptions of the number of CNOTs required to execute our algorithm for each specific case. This approach is the most intuitive and effective way to understand the mechanism of our algorithm and to handle the problem of preparing an equal-superposition state in the simplest manner. The result of our analysis is presented in Table~\ref{tab:N_CNOTs}. The first case corresponds to situations where the desired superposition state can be generated solely using Hadamard gates, i.e., $\hat{H}^{\otimes n}$. This case is straightforward and uncomplicated. Proceeding with the second case, our superposing algorithm can be implemented using only two types of gates: $\hat{G}$ and zero-c$\hat{H}$. Among the cases involving odd $N$, this is the simplest scenario. The remaining cases necessitate all three types of gates: $\hat{G}$, c$\hat{G}$, and zero-c$\hat{H}$, demanding a higher number of CNOTs. Based on these categorizations, we can conclusively provide the CNOT count for our proposed algorithm as follows:
\begin{result}
Given arbitrary $N$ and $g \geq 2$, the number of CNOT gates required in our superposing algorithm is $g + m - 3$, where $g$ is identified from Eq.~(\ref{eq:odd_M}) and $m=\lceil \log_2{M}\rceil$.  When $g=1$, implementing the CNOT gate is unnecessary.
\label{result:CNOTs}
\end{result}
To confirm this result, let us first characterize the Hamming weight (i.e., the number of $1$'s) of the binary-represented $N$. We can present the proof and analyses of $\mathbf{Result~\ref{result:CNOTs}}$ in a structured and coherent manner as follows:
\begin{itemize}
\item[(i)] When $N$ has a Hamming weight of $1$, the circuit can be designed exclusively with the Hadamard gates, resulting in zero CNOT gates.---{\em CASE I.}
\item[(ii)] If $N$ has a Hamming weight of $2$ and $\xi=0$, $\hat{G}$(p) and zero-c$\hat{H}$ gates are utilized to implement the algorithm. In this case, only zero-c$\hat{H}$ gates are counted as $2$-qubit gates, leading to a CNOT count of $n-1$.---{\em CASE II.}
\item[(iii)] When $N$ has a Hamming weight of $n$, we require $n-2$ c$\hat{G}(p)$ gates and $n-1$ zero-c$\hat{H}$ gates, resulting in a total CNOT count of $2n-3$.---{\em CASE III.}
\item[(iv)] For odd $N$ with a Hamming weight $g$ (excluding the case where $N$ has a Hamming weight of $n$ or 2), the total count is $g+n-3$; specifically, the numbers of c$\hat{G}(p)$ and zero-c$\hat{H}$ gates are $g-2$ and $n-1$, respectively. Here, since $g$ is less than $n-1$, a total CNOT count less than $2n-4$.---{\em CASE IV.}
\item[(v)] When $N$ is even and has a Hamming weight 2 or greater than 2, there is at least one $H$. The CNOT count is thus $g+m-3$ that is less than $2n-5$.---{\em CASE V.}
\end{itemize}
Based on analyses (i) through (v), it is evident that the required numbers of c$\hat{G}(p)$ and zero-c$\hat{H}$ gates are $g-2$ and $m-1$, respectively. Therefore, the CNOT count of our algorithm is $g+m-3$, confirming the  $\mathbf{Result~\ref{result:CNOTs}}$. Notably, CASE III demands the highest count of CNOT gates. Consequently, the total number of CNOT gates employed in our algorithm does not exceed $2n - 3$. This result shows the improvement compared to Ref.~\cite{Yu2017} and is consistent with the very recent result in~Ref.~\cite{Shukla2024}. However, note that our approach is more efficient than as we can implement the c$\hat{G}(p)$ gate using only one CNOT, as in Fig.~\ref{fig:z-cH,CG}(b). To emphasize, such a single-CNOT c$\hat{G}(p)$ gate implementation is possible when the target input is in $\ket{0}$, which is suitable for our algorithm (see~{\em Note added}). In Table~\ref{tab:N}, specific values of $N$ associated with individual cases are provided. For those numbers indicating the analyzed features (indicated by an asterisk in Table~\ref{tab:N}), corresponding quantum circuits are shown in Fig.~\ref{fig:circuit_examples}.

\begin{figure}[t]
	\centering
	\includegraphics[width=0.47\textwidth]{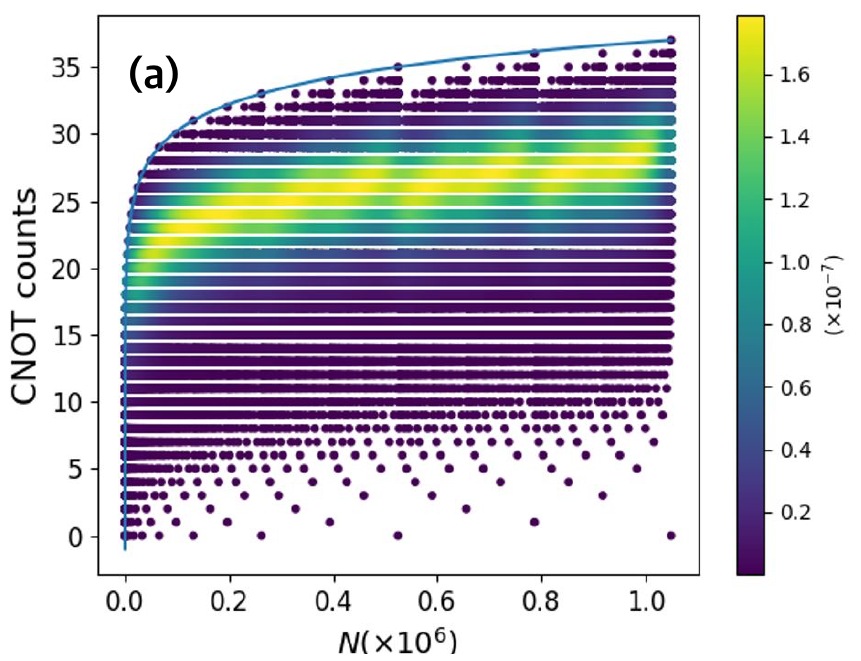}
	\includegraphics[width=0.47\textwidth]{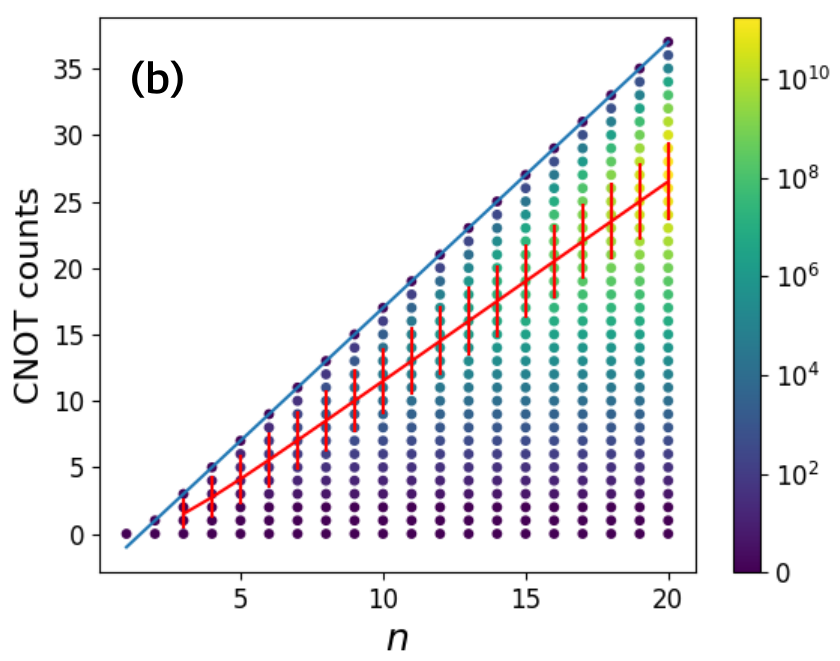}
	\caption{Numerical simulation results up to $20$-qubit scale. In Figure (a), each data point signifies the number of CNOT gates required to generate the quantum superposition of $N$ data. The theoretical analysis, represented by the blue solid line, aligns perfectly with the upper bound data. Different colors are employed to emphasize the density of the data points. In Figure (b), the graph illustrates CNOT counts versus $n$. The maximum of $2n - 3$ CNOT complexity (blue solid line) is clearly confirmed. Additionally, the average CNOT counts are estimated (red solid line), i.e., $\simeq \lceil 1.45n-2.76 \rceil$ (for $n \ge 3$).}
	\label{grp:cnot_count}
\end{figure}

Upon closer analysis, numerical simulations were conducted up to $20$-qubit scale (i.e., $n = 20$, or equivalently, $N = 2^{20}$) using {\bf Algorithm}~\ref{alg1:pse_QS} and~\ref{alg2:pse_QS_odd}. First, we counted the number of CNOT gates by incrementing the value of $N$. The results are illustrated in Fig.~\ref{grp:cnot_count}(a), with distinct colors used to highlight dense regions within the data points. The total count of CNOT gates varies with $N$. For instance, when $N = 2^n$, the parallel applications of $\hat{H}$ are sufficient, and CNOT gates are not used in this case. Conversely, certain $N$ values, e.g., when $N=2^n - 1$, require a significant number of CNOT gates. We plot the graph of $2\log_2{N} - 3$ (represented by the blue solid line), which precisely aligns with the upper bound data. Next, to facilitate a clearer comparison between the theoretical analysis and simulation data, we represent the data as a CNOT count versus $n$ graph [see Fig.~\ref{grp:cnot_count}(b)]. Here, it is confirmed that the upper bound data for CNOT counts precisely aligns with the theoretically derived result of $2n - 3$. Additionally, considering the density of the data points, we estimate the average CNOT counts, which is approximately $\lceil 1.45n-2.76 \rceil$ (for $n \ge 3$).

\section{Discussion}

In this work, we have presented an efficient quantum superposing algorithm for quantum encoding. Within our quantum encoding framework, the efficiency of the quantum superposing algorithm is crucial, specifically considering the classical arbitrary mapping ${\cal M}$ that acts as the bridge connecting classical machine-readable data addresses to the binary indices subsequently encoded into qubits. Such a mapping ${\cal M}$ has been implicitly assumed in studies on computation. Our algorithm significantly enhances computational efficiency for a finite $n$ size of qubit registers by avoiding the use of multi-controlled gates. The improvement in the efficiency of our designed algorithm is due to: (i) the appropriate utilization of the c$\hat{G}(p)$ gate, and (ii) the optimization of the quantum circuit by efficiently implementing c$\hat{G}(p)$ with only one CNOT gate. Through theoretical analysis and numerical simulations, we have demonstrated the effectiveness and superior efficiency of our algorithm, particularly in terms of CNOT complexity. Specifically, our algorithm requires, at most, $2n - 3$ CNOTs, which is the most optimized result so far~\cite{Yu2017,Shukla2024}. This result translates to a substantial improvement in the overall efficiency of quantum encoding.

The advantage of this quantum encoding framework (i.e., the use of an efficient quantum superposing algorithm and mapping ${\cal M}$) lies in the seamless integration of classical and quantum technologies. It allows for a transition from classical to quantum processing when needed, ensuring optimal utilization of classical resources and quantum computational efficiency. This hybrid approach holds significant relevance for utilizing the capabilities of both classical and quantum technologies across a broad spectrum of problems. Another notable aspect of this framework is that during the quantum computation phase, direct access to the actual data is not required. Instead, the mapping ${\cal M}$ can be invoked to identify the final solution after the completion of the main quantum computation. This abstraction layer between data processing and quantum computing contributes to maintaining data integrity and security.

Our study highlights a substantial improvement in the computational efficiency of the quantum superposing algorithm designed for quantum encoding. The advantages of the introduced framework will contribute to the development of a structured approach, seamlessly integrating classical and quantum technologies in quantum encoding. 

{\em Note added.}---We were aware of the publication of Ref.~\cite{Shukla2024} during the review of our work, which report $O(n)$ CNOT counts in preparing the uniform quantum superposition state. This result is consistent with ours, but the algorithm proposed in~\cite{Shukla2024} has the maximum $3n-5$ CNOT counts, whereas our algorithm achieves $2n-3$. This difference arises from the method of decomposing the correlation gates as well as the structure and operation of the algorithm: specifically, we devised a method to implement the c$\hat{G}(p)$ gate using only one CNOT.

\section*{Acknowledgement}
J.B. thanks Prof. M. Choi for discussions and comments. This work was supported by the Ministry of Science, ICT and Future Planning (MSIP) by the Institute of Information and Communications Technology Planning and Evaluation grant funded by the Korean government (2019-0-00003, ``Research and Development of Core Technologies for Programming, Running, Implementing and Validating of Fault-Tolerant Quantum Computing System'') and the National Research Foundation of Korea (NRF-2021M3E4A1038213, NRF-2022M3E4A1077094, NRF-2022M3H3A106307411, NRF-2023M3K5A1094805, and NRF-2023M3K5A1094813).

\section*{References}

\bibliographystyle{iop}

\begin{thebibliography}{10}

\bibitem{Montanaro2016}
Montanaro A 2016 {\it npj Quantum Information\/} {\bf 2} 1

\bibitem{Aaronson2015}
Aaronson S 2015 {\it Nature Physics\/} {\bf 11} 291

\bibitem{Tang2021}
Tang E 2021 {\it Physical Review Letters\/} {\bf 127} 060503

\bibitem{Cotler2021}
Cotler J, Huang H~Y and McClean J~R 2021 {\it arXiv preprint
  arXiv:2112.00811\/}

\bibitem{Weigold2021}
Weigold M, Barzen J, Leymann F and Salm M 2021 {\it IET Quantum
  Communication\/} {\bf 2} 141

\bibitem{Araujo2021}
Araujo I~F, Park D~K, Petruccione F and da~Silva A~J 2021 {\it Scientific
  reports\/} {\bf 11} 6329

\bibitem{Mozafari2021}
Mozafari F, Riener H, Soeken M and De~Micheli G 2021 {\it IEEE Transactions on
  Quantum Engineering\/} {\bf 2} 1

\bibitem{Zhang2022}
Zhang X~M, Li T and Yuan X 2022 {\it Physical Review Letters\/} {\bf 129}
  230504

\bibitem{Biamonte2017}
Biamonte J, Wittek P, Pancotti N, Rebentrost P, Wiebe N and Lloyd S 2017 {\it
  Nature\/} {\bf 549} 195

\bibitem{Bang2019}
Bang J, Dutta A, Lee S~W and Kim J 2019 {\it Physical Review A\/} {\bf 99}
  012326

\bibitem{Havlivcek2019}
Havl{\'\i}{\v{c}}ek V, C{\'o}rcoles A~D, Temme K, Harrow A~W, Kandala A, Chow
  J~M and Gambetta J~M 2019 {\it Nature\/} {\bf 567} 209

\bibitem{Zoufal2019}
Zoufal C, Lucchi A and Woerner S 2019 {\it npj Quantum Information\/} {\bf 5}
  103

\bibitem{Cerezo2022}
Cerezo M, Verdon G, Huang H~Y, Cincio L and Coles P~J 2022 {\it Nature
  Computational Science\/} {\bf 2} 567

\bibitem{Blakeley1996}
Blakeley J~A 1996 in {\it Proceedings of the 1996 ACM SIGMOD international
  conference on Management of data\/} pp. 161--172

\bibitem{Mozafari2022}
Mozafari F, De~Micheli G and Yang Y 2022 {\it Physical Review A\/} {\bf 106}
  022617

\bibitem{De2022}
de~Veras T~M, da~Silva L~D and da~Silva A~J 2022 {\it Quantum Information
  Processing\/} {\bf 21} 204

\bibitem{Araujo2023}
Araujo I~F, Blank C, Ara{\'u}jo I~C and da~Silva A~J 2023 {\it IEEE
  Transactions on Computer-Aided Design of Integrated Circuits and Systems\/}
  1

\bibitem{Shukla2024-2}
Shukla A and Vedula P 2024 {\it arXiv preprint arXiv:2406.13785\/}

\bibitem{Plesch2011}
Plesch M and Brukner {\v{C}} 2011 {\it Physical Review A\/} {\bf 83} 032302

\bibitem{Grover1997}
Grover L~K 1997 {\it Physical review letters\/} {\bf 79} 325

\bibitem{Brassard2002}
Brassard G, Hoyer P, Mosca M and Tapp A 2002 {\it Contemporary Mathematics\/}
  {\bf 305} 53

\bibitem{Viamontes2005}
Viamontes G~F, Markov I~L and Hayes J~P 2005 {\it Computing in science \&
  engineering\/} {\bf 7} 62

\bibitem{Broda2016}
Broda B 2016 {\it The European Physical Journal Plus\/} {\bf 131} 38

\bibitem{Babbush2018}
Babbush R, Gidney C, Berry D~W, Wiebe N, McClean J, Paler A, Fowler A and Neven
  H 2018 {\it Physical Review X\/} {\bf 8} 041015

\bibitem{Yu2017}
Yu Q, Zhang Y, Li J, Wang H, Peng X and Du J 2017 {\it Science China Physics,
  Mechanics \& Astronomy\/} {\bf 60} 070313

\bibitem{Shukla2024}
Shukla A and Vedula P 2024 {\it Quantum Information Processing\/} {\bf 23} 38

\bibitem{Mottonen2005}
M{\"o}tt{\"o}nen M, Vartiainen J~J, Bergholm V and Salomaa M~M 2005 {\it
  Quantum Information \& Computation\/} {\bf 5} 467

\bibitem{Schuld2015}
Schuld M, Sinayskiy I and Petruccione F 2015 {\it Contemporary Physics\/} {\bf
  56} 172

\bibitem{Schuld2018book}
Schuld M and Petruccione F 2018 {\it Supervised learning with quantum
  computers\/} vol.~17 (Springer)

\bibitem{Schuld2021}
Schuld M, Sweke R and Meyer J~J 2021 {\it Physical Review A\/} {\bf 103} 032430

\bibitem{Song2021a}
Song W, Wie{\'s}niak M, Liu N, Paw{\l}owski M, Lee J, Kim J and Bang J 2021
  {\it Quantum Information Processing\/} {\bf 20} 275

\bibitem{Song2021b}
Song W, Lim Y, Kwon H, Adesso G, Wie{\'s}niak M, Paw{\l}owski M, Kim J and Bang
  J 2021 {\it Physical Review A\/} {\bf 103} 042409

\bibitem{Harney2022}
Harney C and Pirandola S 2022 {\it PRX Quantum\/} {\bf 3} 010311

\bibitem{Yoo2014}
Yoo S, Bang J, Lee C and Lee J 2014 {\it New Journal of Physics\/} {\bf 16}
  103014

\bibitem{Lee2019}
Lee J~S, Bang J, Hong S, Lee C, Seol K~H, Lee J and Lee K~G 2019 {\it Physical
  Review A\/} {\bf 99} 012313

\bibitem{Shende2004}
Shende V~V, Markov I~L and Bullock S~S 2004 {\it Physical Review A\/} {\bf 69}
  062321

\bibitem{Shende2005}
Shende V~V, Bullock S~S and Markov I~L 2005 in {\it Proceedings of the 2005
  Asia and South Pacific Design Automation Conference\/} pp. 272--275

\end{thebibliography}

\end{document}